\documentclass[aps,pra,showpacs]{revtex4}

\usepackage{amssymb}
\usepackage{epsfig}
\usepackage{graphicx}

\begin{document}

\preprint{}

\title{Solitons in Bose-Einstein Condensates with time-dependent atomic
scattering length in an expulsive parabolic and complex potential}
\author{Biao Li$^{1,5}$, Xiao-Fei Zhang$^{2,4}$, Yu-Qi Li$^{1,5}$, Yong Chen$^{1,3}$, W. M. Liu$^{2}$}
\affiliation{ $^{1}$Nonlinear Science Center, Ningbo University, Ningbo 315211, China\\
$^{2}$ Institute of Physics, Chinese Academy of Sciences, Beijing 100080, China\\
$^{3}$Institute of Theoretical Computing, East China Normal University, Shanghai, 200062, China\\
$^{4}$Department of Physics, Honghe University, Mengzi 661100, China\\
$^{5}$Key Laboratory of Mathematics Mechanization, Chinese Academy
of Sciences, Beijing 100080, China}

\date{\today}

\begin{abstract}
We present two families of analytical solutions of the
one-dimensional nonlinear Schr\"{o}dinger equation which describe
the dynamics of bright and dark solitons in Bose-Einstein
condensates (BECs) with the time-dependent interatomic interaction
in an expulsive parabolic and complex potential. We also demonstrate
that the lifetime of both a bright soliton and a dark soliton in
BECs can be extended by reducing both the ratio of the axial
oscillation frequency to radial oscillation frequency and the loss
of atoms. It is interested that a train of bright solitons may be
excited with a strong enough background. An experimental protocol is
further designed for observing this phenomenon.
\end{abstract}


\maketitle

\section{Introduction}
The Bose-Einstein condensates (BECs) at nK temperature can be
described by the mean field theory -- nonlinear Schr\"{o}dinger
(NLS) equation with a trap potential, i.e., the Gross-Pitaevskii
(GP) equation. Recently, with the experimental observation and
theoretical studies of BECs \cite{rmp71(1999)463}, there has been
intense interest in the nonlinear excitations of ultra-cold atoms,
such as dark solitons
\cite{prl83(1999)5198,science287(2000)97,prl84(2000)2298,prl85(2000)1598,prl86(2001)2926,
bwu-prl98(2002)034101}, bright solitons
\cite{Nature417(2002)150,prl90(2003)230401,Science296(2002)1290},
vortices \cite{prl83(1999)2498} and the four-wave mixing
\cite{Nature398(1999)218}. Recent experiments have demonstrated
that the variation of the effective scattering length, even
including its sign, can be achieved by utilizing the so-called
Feshbach resonance
\cite{prl82(1999)2422,prl81(1998)5109,nature392(1998)151}. It has
been demonstrated that the variation of nonlinearity of the GP
equation via Feshbach resonance provides a powerful tool for
controlling the generation of bright and dark soliton trains
starting from periodic waves \cite{prl90(2003)230402}.

At the mean-field level, the GP equation governs the evolution of
the macroscopic wave function of BECs. In the physically important
case of the cigar-shaped BECs, it is reasonable to reduce the GP
equation into a one-dimension nonlinear Schr\"{o}dinger equation
with time-dependent atomic scattering length in an expulsive
parabolic and complex potential
\cite{pra57(1998)3837,mplb18(2004)173,prl94(2005)050402,
prl92(2004)220403,mplb18(2004)627,jpb39(2006)3679},
\begin{equation}\label{c01}
{\rm i}\frac{\partial \psi}{\partial t}=
-\frac{\partial^2\psi}{\partial x^2} +2a(t)|\psi|^2\psi
-\frac{1}{4}\lambda^2x^2\psi+{\rm i}\gamma\psi, 
\end{equation}
where the time $t$ and coordinate $x$ are measured in units
$2/\omega_{\perp}$ and $a_{\perp}$, $a_{\perp}=
\sqrt{\hbar/m\omega_{\perp}}$ and $a_{0}=\sqrt{\hbar/m\omega_{0}}$
are linear oscillator lengths in the transverse and cigar-axis
directions, respectively. $\omega_{\perp}$ is the radial oscillation
frequency and $\omega_0$ is the axial oscillation frequency. $m$ is
the atomic mass, $|\lambda|= 2|\omega_0|/ \omega_{\perp}\ll 1$,
$a(t)$ is a scattering length of attractive interactions ($a(t)<0$)
or repulsive interactions ($a(t)>0$) between atoms, and $\gamma$ is
a small parameter related to the feeding of condensate from the
thermal cloud \cite{prl81(1998)933}. When $a(t)=g_0\exp(\lambda t)$
and $\gamma=0$, Liang et al. present a family of exact solutions of
(\ref{c01}) by  Darboux transformation and analyze the dynamics of a
bright soliton \cite{prl94(2005)050402}. Kengne et al. investigated
(\ref{c01}) with  $a(t)=g_0$ and $\gamma=\lambda/2$ and verified the
dynamics of a bright soliton proposed \cite{jpb39(2006)3679}. These
results show that, under a safe range of parameters, the bright
soliton can be compressed into very high local matter densities by
increasing the absolute value of the atomic scattering length or
feeding parameter.

In this paper, we develop a direct method to derive two families
of exact solitons of Eq. (\ref{c01}), then give some thorough
analysis for a bright soliton, a train of bright solitons and a
dark soliton. Our results show that for BEC system with
time-dependent atomic scattering length, the lifetime of a bright
or a dark soliton in BECs can keep longer times by reducing both
the ratio of the axial oscillation frequency to radial oscillation
frequency and the loss of atoms. It is demonstrated that a train
of bright solitons in BECs may be excited with a strong enough
background. We also propose an experimental protocol to observe
this phenomenon in further experiments.

\section{The Method and Soliton Solutions}

We can assume the solutions of Eq. (\ref{c01}) as follows
\begin{equation}\label{c02}
\psi = [ A_{{0}} ( t ) +A_{{1}} ( t){\frac { \delta\,\cosh(\xi)
+\cos(\eta) }{\cosh(\xi)+\delta\,\cos(\eta) }}  +{{\rm i} B_{{1}} (
t )\frac { \alpha\,\sinh(\xi) +\beta\,\sin(\eta) }{ \cosh(\xi)
+\delta\,\cos(\eta) }} ]\exp({\rm i}\Delta),
\end{equation}
where
$$
\Delta = k_{{0}}(t) +k_{{1}}(t) x +k_{{2}}(t){x}^{2},\ \ \xi
=p_{{1}}(t) x +q_{{1}}(t),\ \ \eta = p_{{2}}(t) x+q_{{2}}(t),
$$
and $ A_0(t)$, $A_1(t)$, $B_1(t)$, $p_1(t)$, $q_1(t)$, $p_2(t)$,
$q_2(t)$, $k_0(t)$, $k_1(t)$, $k_2(t)$ are real functions of $t$ to
be determined, and $\alpha, \beta, \delta$ are real constants.

Substituting Eq. (\ref{c02}) into  Eq. (\ref{c01}), we first remove
the exponential terms, then collect coefficients of $\sinh^i(\xi)
\cosh^j(\xi) \sin^m(\eta) \cos^n(\eta) x^k$ ($i=0, 1, 2,\cdots$;
$j=0, 1$; $m=0, 1, 2, \cdots$; $n=0,1$; $k=0, 1, \cdots.$) and
separate real part and imaginary part for each coefficient. We
derive a set of ordinary differential equations (ODEs) with respect
to $a(t)$, $A_0(t)$, $A_1(t)$, $B_1(t)$, $p_1(t)$, $q_1(t)$,
$p_2(t)$, $q_2(t)$, $k_0(t)$, $k_1(t)$, $k_2(t)$. Finally, solving
these ODEs, we can obtain two families of analytical solutions of
Eq. (\ref{c01}).

{\bf Family 1.} When interaction between atoms is attractive such as
$^7$Li atoms, $a(t)<0$, the solution of Eq. (1) can be written as:
\begin{equation}\label{c03}
\psi_1  =  \Omega [ A_c+A_s{\frac {
( \delta\,\cosh(\xi)
+\cos(\eta)) }{\cosh(\xi) +\delta\,\cos(\eta) }}+{\rm i}A_s{\frac {
\alpha\,\sinh(\xi) +\beta\,\sin(\eta) }{ \cosh(\xi)
+\delta\,\cos(\eta) }} ]\exp({\rm i}\Delta+\gamma t),
\end{equation}
where
\begin{eqnarray*}
\Delta &=& k_2(t)x^2+k_1\Omega^2x +(2g_0A_c^2-k_1^2)\int\!\!\Omega^4{\rm d}t,\\
\xi &=& \!\!\sqrt{g_0} A_s[\beta \Omega^2x-\!\!
             2[k_1\beta+p_2\!\! -\frac{2g_0p_2A_c^2}{(g_0A_s^2
             +p_2^2)}]\!\! \int\!\!\Omega^4{\rm d}t],\\
\eta &=& p_2\Omega^2x-[2p_2k_1 +(p_2^2-g_0A_s^2)\beta]\int\!\Omega^4{\rm d}t,\\
a(t) &=& -g_0\Omega^2\exp(-2\gamma t),\ \ \alpha =
            -\frac{2\sqrt{g_0}A_cp_2}{g_0A_s^2+p_2^2},\\
\beta^2 &=& \frac{p_2^2 +g_0(A_s^2
            -4A_c^2)}{g_0A_s^2+p_2^2},\ \ \delta=-\frac{2g_0A_cA_s}{g_0A_s^2+p_2^2},\\
\Omega &=& {\rm exp}[\int \!-2k_{2}(t){\rm d}t],\ \
              k_{{2}}(t) =\{\pm\frac{\lambda}{4},\ \ \frac{\lambda}{4}{\tanh(\lambda t)}\},
\end{eqnarray*}
and $A_c, A_s, g_0>0$, $p_2$, $k_1$, $\gamma$ are arbitrary real
constants.

{\bf Family 2.} When interaction between atoms is repulsive such as
$^{23}$Na and $^{87}$Rb atoms, $a(t)>0$, the solution of Eq. (1) can
be written as:
\begin{equation}\label{c04}
\psi_2=\Omega [A_c+iA_s\tanh(\xi)]\exp(i\Delta+\gamma
t),
\end{equation}
where 
\begin{eqnarray*}
 \xi &=& \pm\sqrt{g_0}\Omega^2 x +2A_s(\sqrt{g_0}k_1
               +g_0A_c)\int\!\Omega^4 {\rm d}t,\\
\Delta &=& k_2(t)x^2\!+\!k_1\Omega^2x
             \!-\![2g_0(A_c^2\!+\!A_s^2)\!+\!k_1^2]\int\!\! \Omega^4{\rm d}t,\\
a(t)&=& g_0\Omega^2\exp(-2\gamma t),
\end{eqnarray*}
and $\Omega, k_2(t)$ are the same as in (\ref{c03}), $A_c, A_s,
g_0>0$, $k_1$ and $\gamma$ are arbitrary real constants.

The solutions (\ref{c03}) and (\ref{c04}) are new general solutions
of equation (\ref{c01}) which can describe the dynamics of bright
and dark solitons in BECs with the time-dependent interatomic
interaction in an expulsive parabolic and complex potential. In
special case, it can be reduced to solutions obtained by others. For
example, if $k_2(t)=-\lambda/4$ and $\gamma=0$, the solution
(\ref{c03}) describe dynamics of a bright soliton in BECs with
time-dependent atomic scattering length in an expulsive parabolic
potential, and it can reduce the solution in Ref.
\cite{prl94(2005)050402}. If $k_2(t)=-\lambda/4$ and
$\gamma=\lambda/2$, Eq. (\ref{c03}) describe dynamics of bright
matter wave solitons in BECs in an expulsive parabolic and complex
potential, and it can recover the solution in Ref.
\cite{jpb39(2006)3679}.

To our knowledge, the other solutions from Eqs. (\ref{c03}) and
(\ref{c04}) have not been reported earlier. When
$k_2(t)=\pm\lambda/4$ and $\gamma$ is a fixed value, the intensities
of Eqs. (\ref{c03}) and (\ref{c04}) are either exponentially
increasing or exponentially decreasing  so the BECs phenomenon can
not be stable reasonably. Thus in order to close experimental
condition and compare our theoretical prediction with experimental
results, we will only discuss and analyze Eqs. (\ref{c03}) and
(\ref{c04}) with $k_2(t)=\lambda/4\tanh(\lambda t)$ and
$\Omega^2={\rm sech}(\lambda t)/2$.

\subsection{Dynamics of Bright Solitons in BECs}
 In the following, we are interested in two cases of Eq. (\ref{c03}).

{\bf (I)}\ \ When $\alpha=0$ and $p_2=0$, $\psi_1$ can be written as
\begin{equation}\label{c05}
    \psi_{11}= \Omega [ A_c+A_s\frac{\delta\cosh(\xi)
                     +\cos(\eta)+{\rm i} \beta\,\sin(\eta)}{\cosh(\xi) +\delta\,\cos(\eta)}]  \times\exp({\rm i}\Delta+\gamma t),
\end{equation}
where  $A_s^2>4A_c^2$, and
\begin{eqnarray*}
  \xi &=& \sqrt{g_0}\beta A_s\Omega^2x-2\sqrt{g_0}A_s k_1\beta\int\!\!\Omega^4{\rm d}t, \\
  \eta  &=& \frac{g_0(A_s^2-4A_c^2)}{\beta}\int\!\Omega^4{\rm d}t,\ \ \beta^2=\frac{A_s^2-4A_c^2}{A_s^2}, \\
  \Delta &=& k_2(t)x^2+k_1\Omega^2x+(2g_0A_c^2-k_1^2)\int\Omega^4{\rm d}t,\\
 \Omega &=& \sqrt{\frac{{\rm sech}(\lambda t)}{2}},\ \ \delta =
 -\frac{2A_c}{A_s},\\
 a(t)&=&-\frac{g_0}{2}\rm{sech}(\lambda t) {\rm exp}(-2\gamma t).\\
\end{eqnarray*}
When $A_s=0$, $\psi_{11}$ reduces to the background
  \begin{equation}\label{c07}
  \psi_{c}=A_c\Omega\exp({\rm
i}\Delta+\gamma t).
\end{equation}
 When $A_c=0$, $\psi_{11}$
reduces to the bright soliton 
\begin{equation}\label{c06}
\psi_{s}=A_s\Omega {\rm sech}(\xi)\exp\left[{\rm i}\Theta+\gamma
t\right],
\end{equation}
where
\begin{eqnarray*}
 \xi &=& \sqrt{g_0} A_s\Omega^2x -2\sqrt{g_0}A_s(k_1+p_2)\int\Omega^4{\rm d}t,\\
\Theta &=& k_2(t)x^2+k_1\Omega^2x +[g_0A_s^2-k_1^2]\int\Omega^4{\rm
d}t.
\end{eqnarray*}

Thus $\psi_{11}$ represents a bright soliton embedded in the
background. At the same time, when $A_c\ll A_s$ satisfied
$4A_c^2<A_s^2$ and $\gamma$ small, the background is small within
the existence of bright soliton.

Considering the dynamics of the bright soliton in the background,
the length $2L$ of the spatial background must be very large
compared to the scale of the soliton. In the real experiment
\cite{Nature417(2002)150}, the length of the background of BECs
can reach at least $2L=370{\rm \mu m}$. In Fig.1, the width of the
bright soliton is about $2l=14{\rm nm}$ [a unity of coordinate,
$\Delta x=1$ in the dimensionless variables, corresponds to
$a_{\perp}=(\hbar/(m\omega_{\perp})^{1/2}=1.4{\rm \mu m}$]. So
$l\ll L$, a necessary condition for realizing bright soliton in
experiment. From Fig.1a, under the realistic experiment parameters
in \cite{Science296(2002)1290}, i.e., $\omega_{\perp}=2\pi\times
710$Hz, $\omega_0=2{\rm i}\pi\times 70$Hz. In order to cope with
the experiment: the soliton move to $-x$ direction,
$\lambda=-2|\omega_0|/\omega_{\perp}\approx -0.197$,
$\gamma=-0.01$ and $g_0=0.4$nm , we can see that the lifetime of
the BEC is about $20\times 4.5\times 10^{-4}s=9$ms, which is close
to the experiment results: the lifetime of a BEC is about 8ms.
Here, by (\ref{c09}), we can verify that the number of atoms in
the bright soliton against the background is in the range of 4635
(at $t=-10$) and 3107 (at $t=10$), which is a proper range of
atoms when the soliton can be observed
\cite{Science296(2002)1290}. But, when $t\in [-10,10]$, the
scattering length $a(t)=-0.2{\rm sech}(0.197t)\exp(0.02t)$ varies
in [$-0.20, -0.04$] {\rm nm}, which is different from the
experiment condition: the scattering length keeps invariable when
the bright soliton in BECs propagates in the magnetic trap
\cite{Science296(2002)1290}. However, up to now, they can not
measure the motion of dark or bright soliton when the parameter
varies continually, and they only measure a particular value of
soliton corresponding to the fixed magnetic field. When we fix the
magnetic field at a fixed value (the scattering length is also a
fixed value) such as $B=425$G in Ref. \cite{Science296(2002)1290},
the scattering length is $a_s = -0.21$ nm, then the special value
of our general solution is in accordance with the experimental
data of Ref. \cite{Science296(2002)1290}. Of course, we believe
that with the development of the Feshbach resonance technology,
the experimental physicists can measure the motion of solitons in
the future. Thus by modulating the scattering length in time via
changing magnetic field near the Feshbach resonance, we may also
realize the bright solitons in BECs.

\begin{figure}[!ht]
\centering
\renewcommand{\figurename}{Fig.}
\includegraphics[height=5cm,width=7cm]{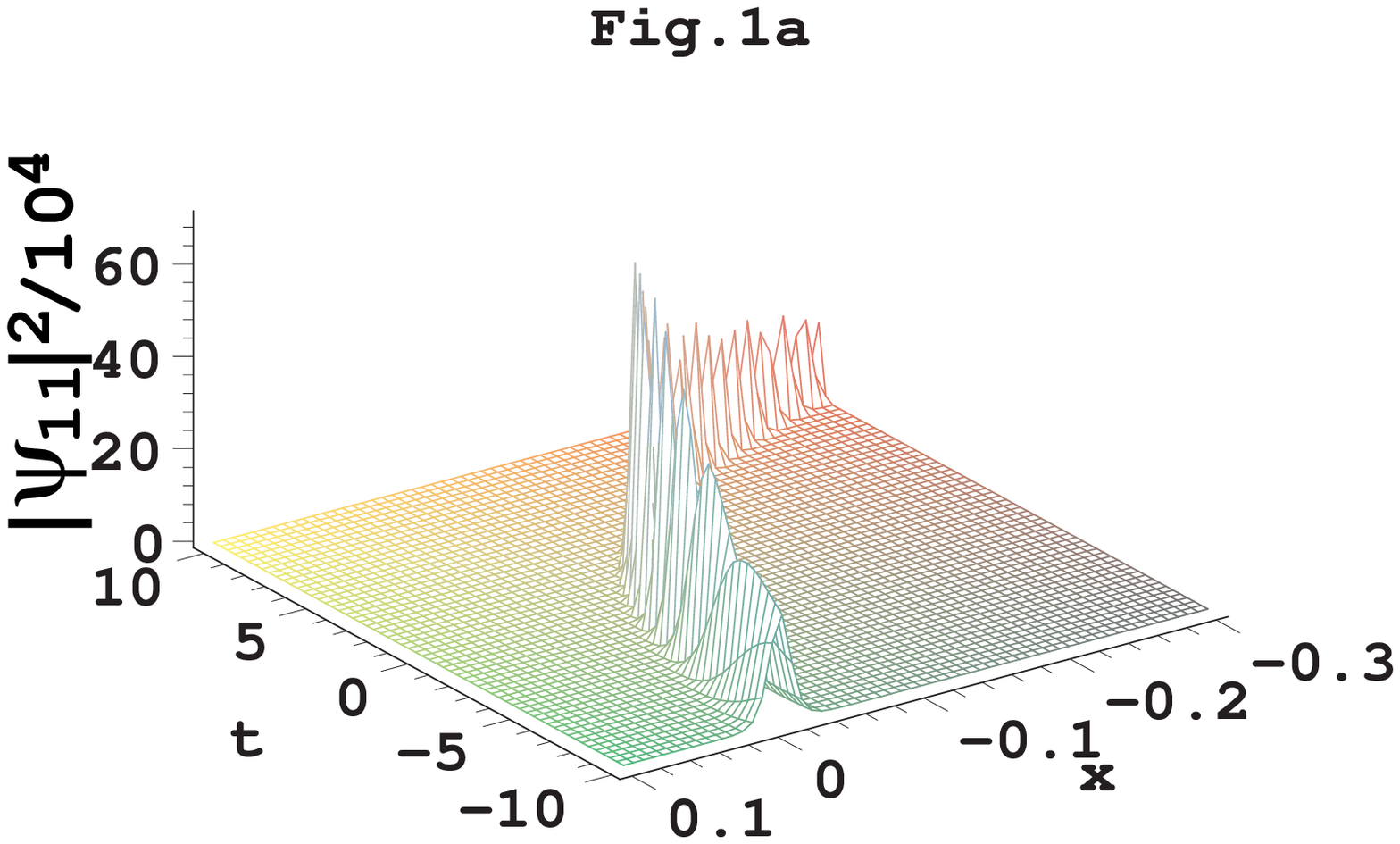}%
\includegraphics[height=5cm,width=7cm]{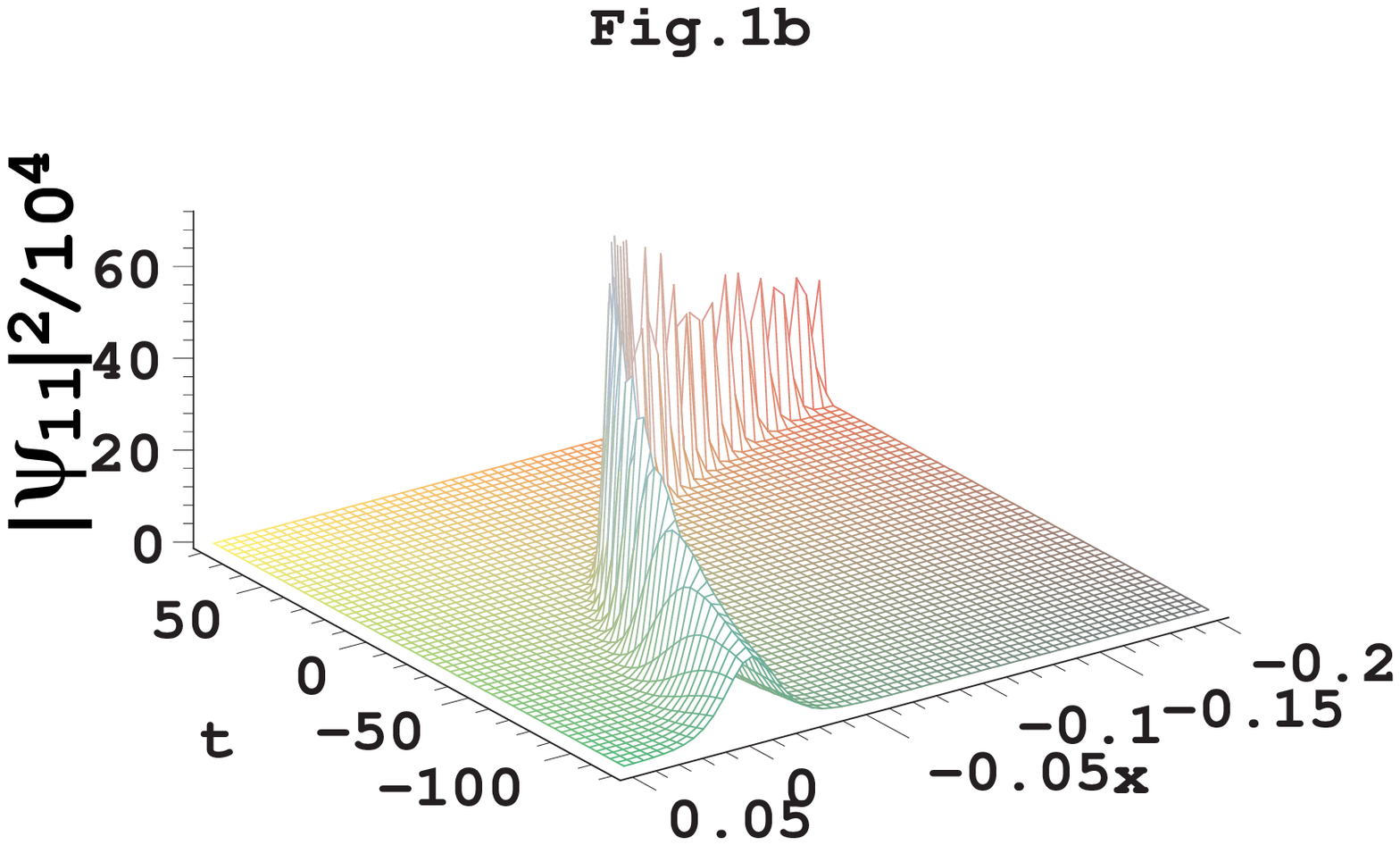}
\caption{\small (Color online) The evolution plots of
$|\psi_{11}|^2/10^4$, where $A_c=4, A_s=1200, g_0=0.4$. In Fig.1a:
$\lambda=-0.197, \gamma=k_1=-0.01$; In Fig.1b: $\lambda=-0.02,
\gamma=k_1=-0.001$. }
\end{figure}

When $\lambda=-0.02$ (which can be derived from $\omega_0=2{\rm
i}\pi\times 7$Hz and $\omega_{\perp}=2\pi\times 710$Hz) and
$\gamma=-0.001$, from Fig.1b, the lifetime of the BEC can reach
about 200 unities of the dimensionless time corresponds to a real
time of 0.1s, which arrives at the order of the lifetime of a BEC
in today's experiments. Here, we can verify that when $t$ is from
-120 to 80, (i) the number of atoms is in the range of 4824 and
3234 by (\ref{c09}); (ii) the scattering length $-0.20{\rm nm}\leq
a(t)\leq -0.03{\rm nm}$; (iii) From (5), we can derive $\xi\approx
-1/2\sqrt{g_0}\sqrt{A_s^2-4A_c^2}(x-k_1t-x_0)$, therefore, $k_1$
describes the velocity of the bright soliton, which can be
demonstrated by Fig. 1.

It is necessary to point out that, (i) In order to give clear
figures, the parameter $k_1$ is taken to be relatively small. In
reality, $k_1$ is about to $-4$, which can be derived from the
soliton's position $x=-k_1\exp(-\lambda t)/\lambda$; (ii) The
background is very small with regard to the bright soliton in Fig.
1, which can be also shown by Fig. 2. Therefore the background may
be taken as zero background approximatively; (iii) In order to keep
the lifetime of the bright soliton about {\rm 8ms}, we take
$\gamma=-0.01$ which may be the experimental value. When
$\gamma=-0.01$, by varying the value of $\lambda$, it is difficult
to extend the lifetime of the bright soliton. Thus in order to
extending the lifetime of a soliton in BEC, we should take
appropriate measures to reduce the absolute value of $\lambda$ and
$\gamma$, as is shown in Fig. 1b.


\begin{figure}
\centering
\renewcommand{\figurename}{Fig.}
\includegraphics[height=3cm,width=7cm]{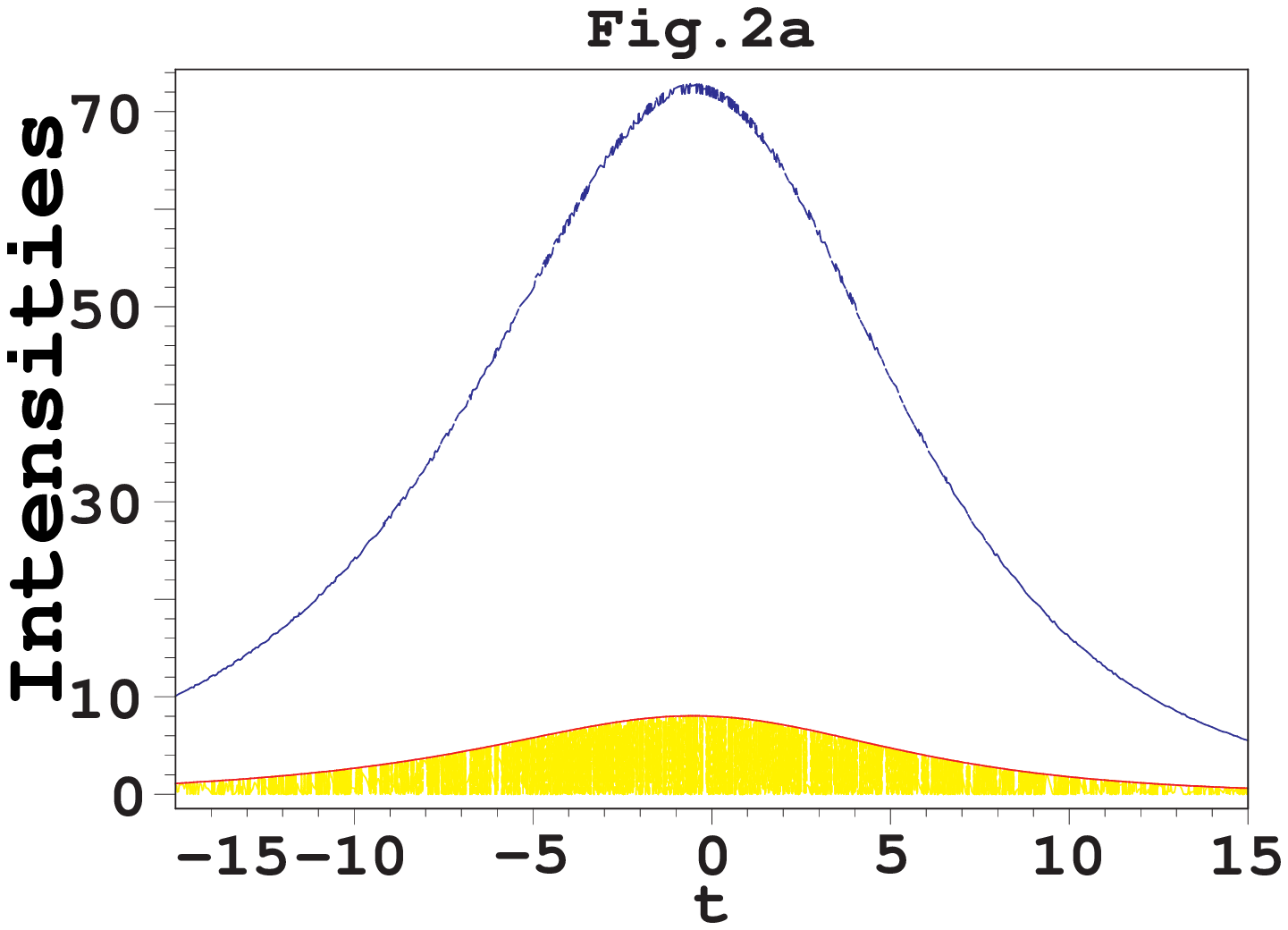}%
\vspace{0.1cm}
\includegraphics[height=3cm,width=7cm]{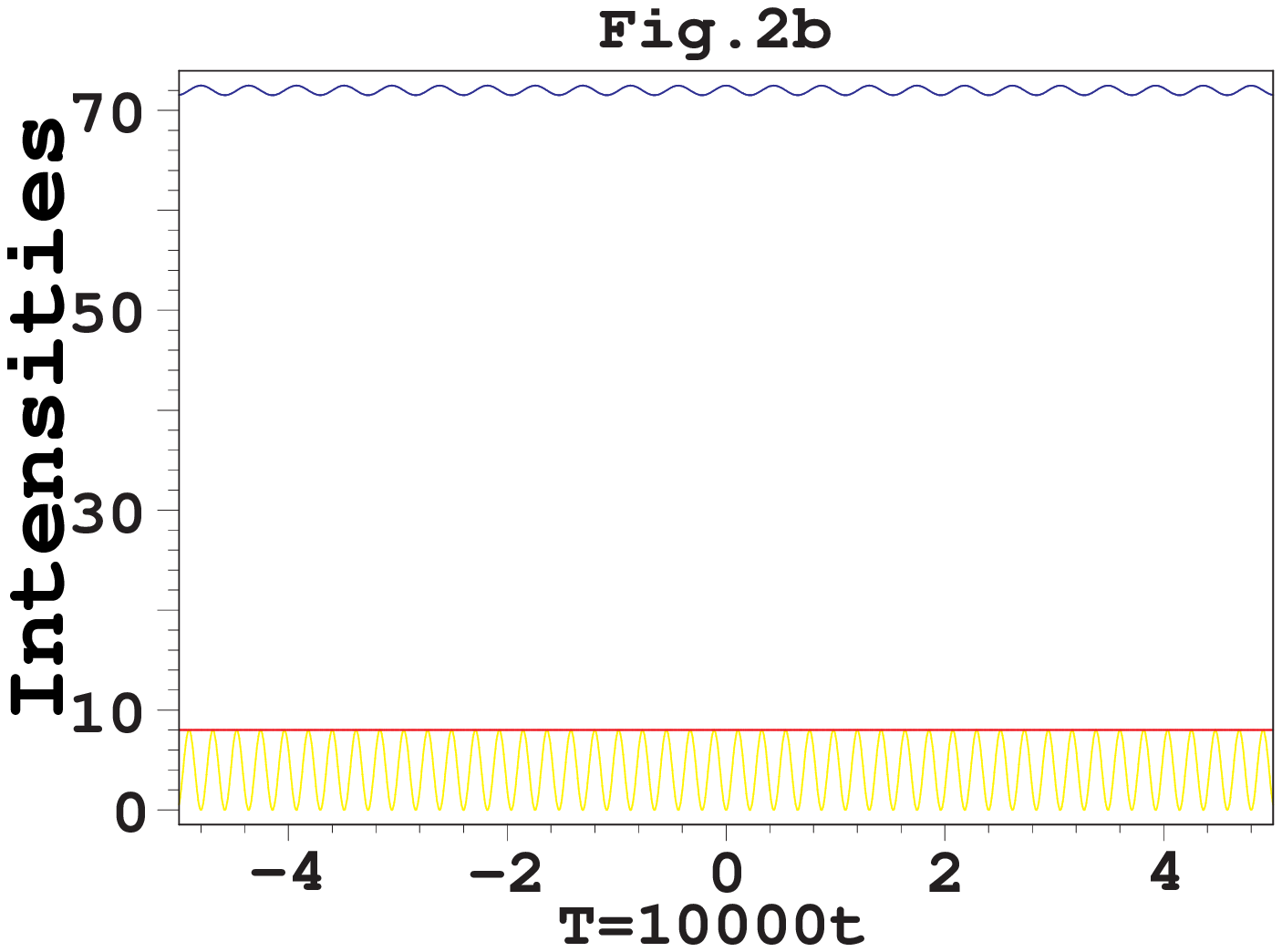}%
\vspace{0.1cm}
\includegraphics[height=3cm,width=7cm]{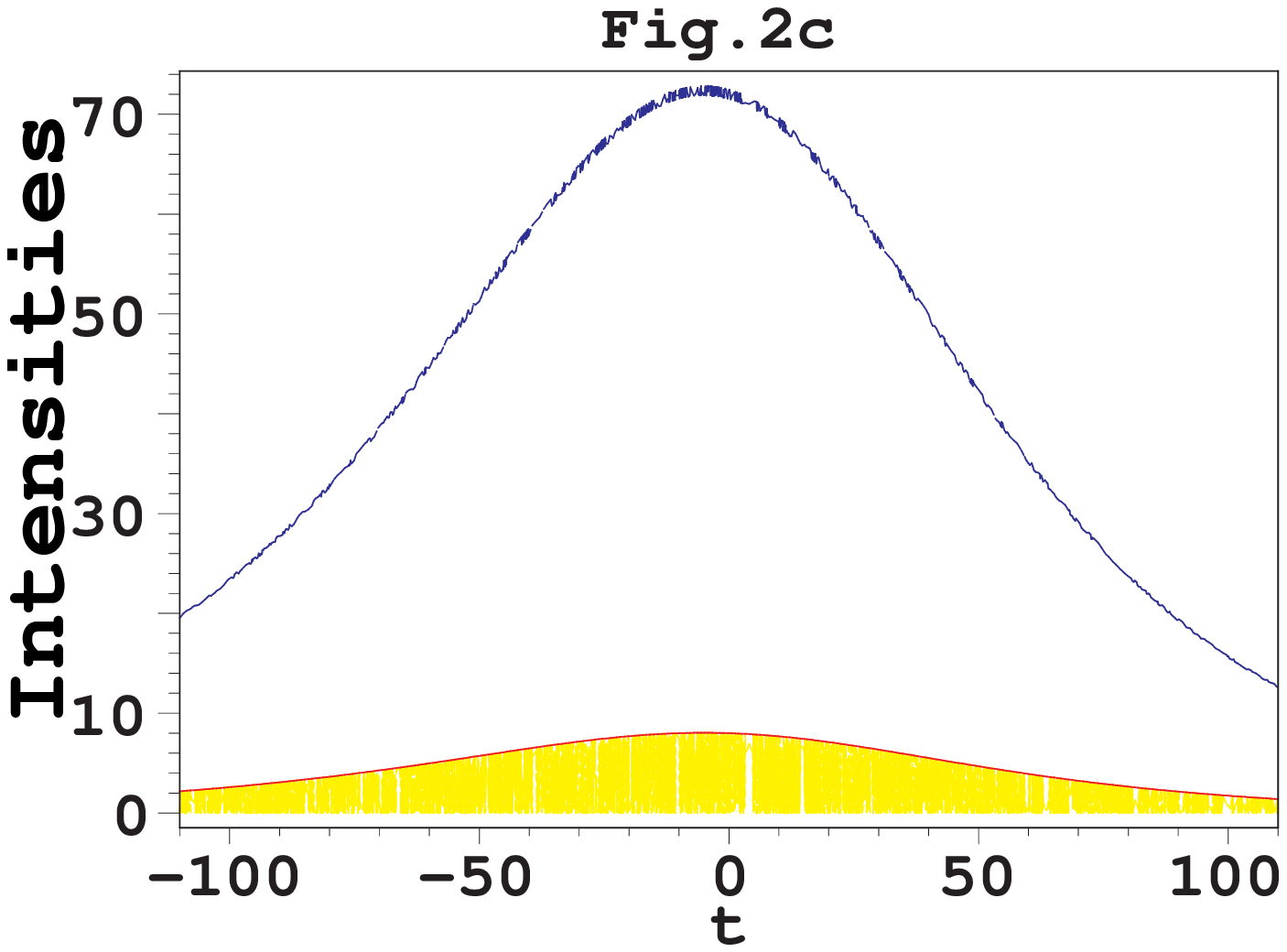}%
\vspace{0.1cm}
\includegraphics[height=3cm,width=7cm]{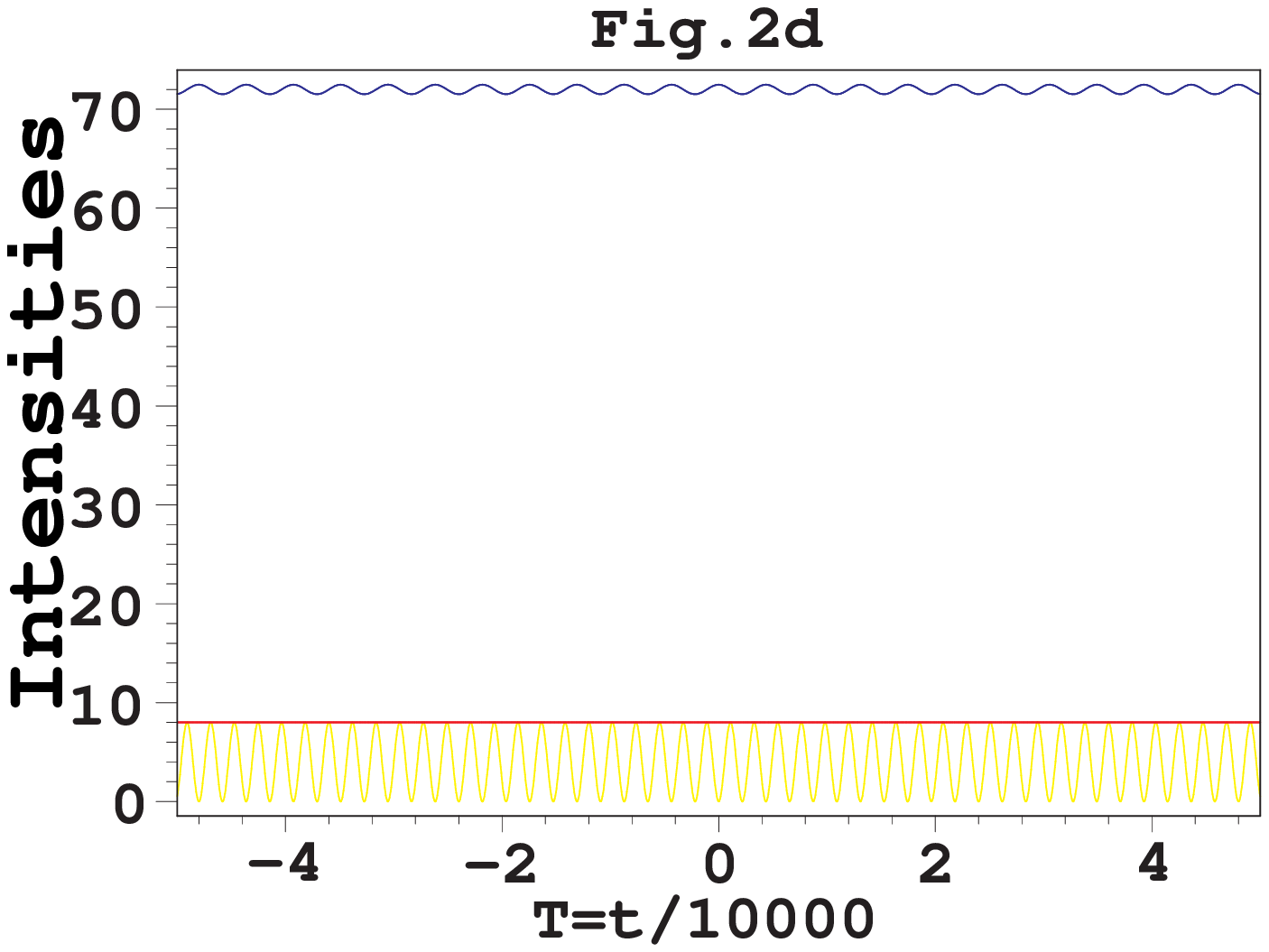}%
\caption{\small (Color online) The evolution plots of
$|\psi_{11}|^2_{\rm max}/10^4$ (blue line) and $|\psi_{11}|^2_{\rm
min}$ (yellow line) and $|\psi_c|^2$ (red line). The parameters in
Fig. 2a-2b and Fig. 2c-2d are the same as in Fig. 1a and Fig. 1b,
respectively.}
\end{figure}

Furthermore, we find that when $\cosh(\xi)=
-2A_c\cos(\eta)/A_s+A_s/(A_c\cos(\eta))$ ($\sinh(\xi)=0$), the
intensity of (\ref{c05}) arrives at the minimum (maximum)
\begin{eqnarray}\label{c08}
 \nonumber |\psi_{11}|^2_{\rm min} &=& A_c^2[1-\frac{(A_s^2-4A_c^2)
\cos^2(\eta)}{A_s^2-4A_c^2\cos^2(\eta)}] \frac{{\rm sech}(\lambda
t)}{2}{\rm e}^{2\gamma t}, \\
   |\psi_{11}|^2_{{\rm max}} &=& [A_c^2+\frac{A_s(A_s^2 -4A_c^2)}{A_s -2A_c\cos(\eta)}] \frac{{\rm
sech}(\lambda t)}{2}{\rm e}^{2\gamma t}.
\end{eqnarray}
This means that the bright solitons (\ref{c05}) can be only squeezed
into the assumed peak matter density between the minimum and maximum
values. Fig.2 present the evolution plots of the maximal and minimal
intensities given by $|\psi_{11}|^2_{\rm max}/10^4$ (blue line)  and
$|\psi_{11}|^2_{\rm min}$ (yellow line) and the background intensity
(red line) with different parameters. From Fig.2, with the time
evolution, firstly the intensities increase until to the peak, then
decrease to the background. Meanwhile, the smaller $|\lambda|$ and
$|\gamma|$, the longer the higher intensities can keep. Therefore in
order to keep a bright soliton in BECs for longer time, we should
reduce both the ratio of the axial oscillation frequency to radial
oscillation frequency and the loss of atoms.

To investigate the stability of the bright soliton in the expulsive
parabolic and complex potential, we obtain 
\begin{equation}\label{c09}
\int_{-L}^{+L}\!\!(|\psi_{11}|^2-|\psi_{c}|^2)dx=N_0C_L,
\end{equation}
where
\begin{eqnarray}\label{c10} 
  N_0 &=& \frac{2\sqrt{A_s^2-4A_c^2}}{\sqrt{g_0}}\exp(2\gamma t) \\
  C_L &=&
  \frac{A_s(\exp(2L)-1)}{A_s(\exp(2L)+1)-4A_c\cos(\eta)\exp(L)},\label{c11}
\end{eqnarray}
which is the exact number of the atoms in the bright soliton against
the background described by (\ref{c05}) within $[-L, L]$. This
indicates that when $L$ takes a fixed value, for example, $L=100$,
then $C_L\rightarrow 1$, therefore the number of atoms in the bright
soliton is determined by $N_0$.

In contrast, the quantity 
\begin{equation}\label{c12}
\int_{-L}^{L}|\psi_{11}-\psi_{c}|^2dx =N_0
\{C_L+4A_cM\cos(\eta)\}, 
\end{equation}
where $N_0$ and $C_l$ are determined by (\ref{c10}) and (\ref{c11}),
and
\begin{eqnarray*}
M &=& \frac{1}{\Theta}\arctan[\frac{A_s+2A_c\cos(\eta)}{\Theta}\cdot\frac{\exp(L)-1}{\exp(L)+1}],\\
\Theta &=& \sqrt{A_s^2-4A_c^2\cos^2(\eta)},
\end{eqnarray*}
counts the number of atoms in both the bright soliton and background
under the condition of $\psi(\pm L,t)\neq 0$. Equation (\ref{c12})
displays that a time-periodic atomic exchange is formed between the
bright soliton and the background. In the case of zero background,
i.e., $A_c=0$, from (\ref{c12}) the exchange of atoms depends on the
sign of $\gamma$: (i) when $\gamma=0$, there will be no exchange of
atoms; (ii) when $\gamma<0$, the exchange of atoms decreases; (iii)
when $\gamma>0$, the exchange of atoms increases. As shown in Fig.
3, in the case of nonzero background and $\gamma<0$, a
slow-fast-slow process of atomic exchange is performed between the
bright soliton and the background, but the whole trend of the atomic
exchange between the bright soliton and the background is decrease.
In Ref. \cite{pra75(2007)037601}, Wu et al show that the number of
atoms continuously injected into Bose-Einstein condensate from the
reservoir depends on the linear gain/loss coefficient, and cannot be
controlled by applying the external magnetic field via Feshbach
resonance. The findings here can recover the same results.

\begin{figure}[!ht]
\centering
\renewcommand{\figurename}{Fig.}
\includegraphics[height=7.5cm,width=7cm]{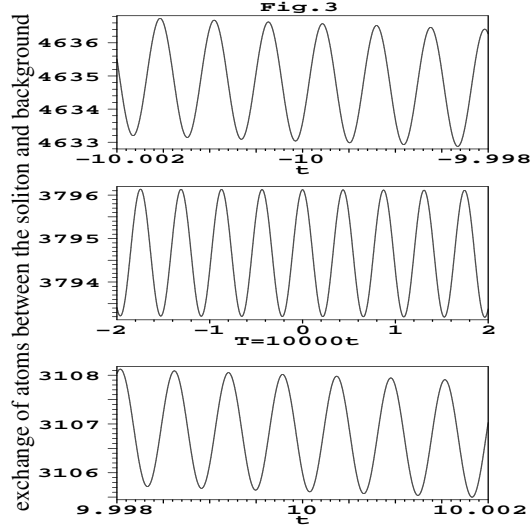}%
\caption{\small (Color online) The atomic exchange between the
bright solitons and the background given by (\ref{c12}) with
$\lambda=-0.197, A_c=4, A_s=1200, g_0=0.4, \gamma=-0.01$. }
\end{figure}

\begin{figure}[!ht]
\centering
\renewcommand{\figurename}{Fig.}
\includegraphics[height=5cm,width=7cm]{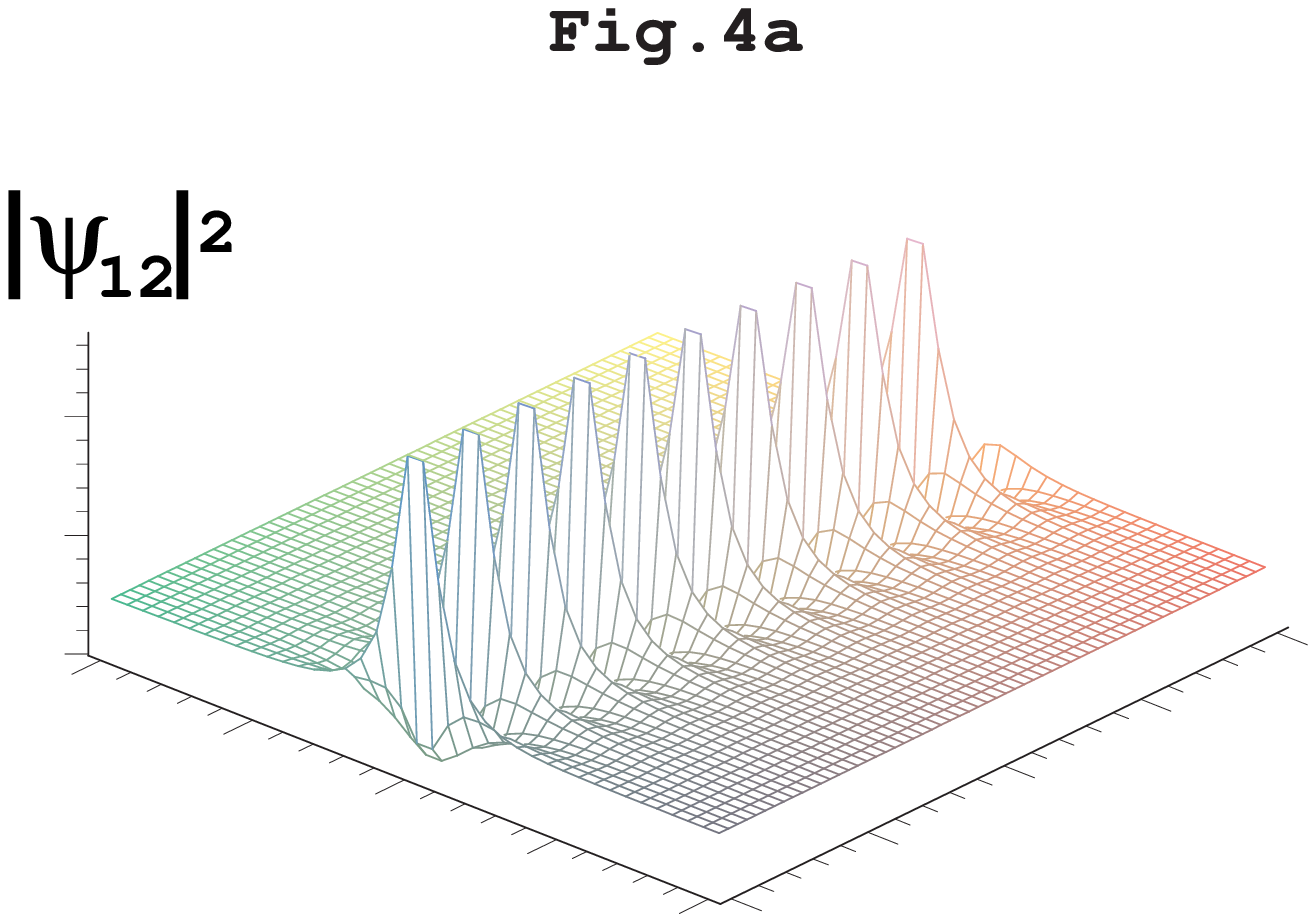}%
\includegraphics[height=5cm,width=7cm]{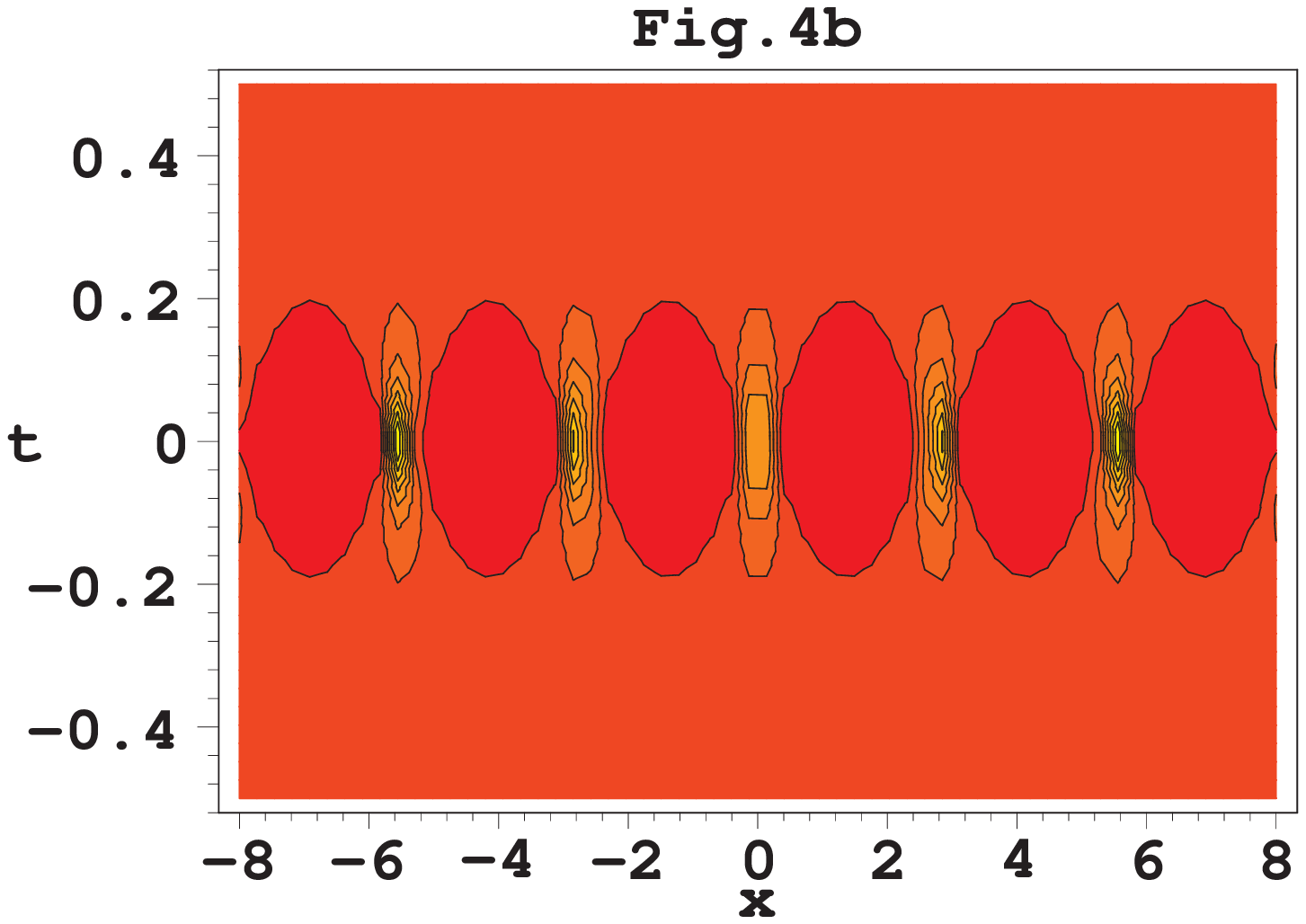}
\caption{\small (Color online) The evolution plots of
$|\psi_{12}|^2$ with $\lambda=-0.197, A_c=7, A_s=12, g_0=0.4,
k_1=-0.01, \gamma=-0.01$. Fig. 4b is the contour plot. }
\end{figure}

In addition, under the integration constant of $\int\Omega^4(t){\rm
d}t$ taken to be zero, (\ref{c05}) take the following particular
form at $t_0=\frac{1}{2\lambda} \ln[-\frac{g_0(A_s^2
-4A_c^2)}{\lambda\beta(2k+1)\pi}-1]$ $(k=0,\pm 1, \pm 2,\cdots,)$.
\begin{equation}\label{c13}
\psi_{11}=\Omega(t_0)\left[-A_c\pm iA_s\beta{\rm
sech}(\xi)\right]\exp(i\Delta+\gamma t_0).
\end{equation}

This means that (\ref{c13}) can be generated by coherently adding a
bright soliton into the background.

Inspired by two experiments \cite{Nature417(2002)150,
Science296(2002)1290}, we can design an experimental protocol to
control the soliton in BECs near Feshbach resonance with the
following steps: (i) Create a bright soliton in BECs with the
parameters of $N\approx 4\times 10^3$, $\omega_{\perp}=2{\rm
i}\pi\times 700$Hz and $\omega_0=2\pi\times 7$ Hz, and for $^7$Li.
(ii) Under the safe range of parameters discussed above, ramp up the
absolute value of the scattering length according to
$a(t)=-\frac{g_0}{2}\rm{sech}(\lambda t) {\rm exp}(-2\gamma t)$ due
to Feshbach resonance, control the dispersion of atoms in BECs at a
low level by modulating the parameter $\gamma$ about to $-0.001$,
and take $\lambda$ to be a very small value:
$\lambda=-2|\omega_0|/\omega_{\perp}=-0.02$. A unity of time,
$\Delta t=1$ in the dimensionless variables, corresponds to real
seconds $2/\omega_{\perp}=4.5\times 10^{-4}$. (iii) During 200
dimensionless units of time, the absolute value of the atomic
scattering length varies in $0.03{\rm nm}\leq |a(t)|\leq 0.20{\rm
nm}$. This means that during the process of the bright soliton, the
stability of soliton and the validity of 1D approximation can be
kept as displayed in Fig. 1b. Therefore, the phenomena discussed in
this paper should be observable within the
current experimental capability.\\

{\bf (II)} When $\beta=0$, the solution $\psi_1$ is written as
\begin{equation}\label{c14}
  \psi_{12} = \Omega [A_c+A_s{\frac{\delta\,\cosh(\xi)
+\cos(\eta)+i \alpha\,\sinh(\xi) }{\cosh(\xi) +\delta\,\cos(\eta)
}}]  \times \exp(i\Delta+\gamma t),
\end{equation}
where $4A_c^2-A_s^2>0$ and
\begin{eqnarray*}
 \xi &=& \sqrt{g_0}A_sp_2\int\Omega^4{\rm d}t,\ \ \alpha =-\frac{p_2}{2\sqrt{g_0}A_c},\\
\eta &=& p_2\Omega^2x -2p_2k_1\int\!\Omega^4{\rm d}t,\ \ \delta=-\frac{A_s}{2A_c},\\
p_2^2 &=& g_0(4A_c^2-A_s^2),\ \ \Omega=\sqrt{\frac{{\rm
sech}(\lambda t)}{2}}.
\end{eqnarray*}
Analysis reveals that $\psi_{12}$ is periodic with a period
$\Gamma=4\pi/[{p_2{\rm sech}(\lambda t)}]$ in the space coordinate
$x$ and aperiodic in the temporal variable $t$. Note that the
period $\Gamma$ is not a constant due to the presence of the
function ${\rm sech}(\lambda t)$, but when $\lambda\ll 1$ and $t$
is very small, $\Gamma$ is very close to $4\pi/p_2$. As shown in
Fig.4, when $\lambda=-0.197$, $A_c=7, A_s=12, g_0=0.4$,
$k_1=\gamma=-0.01$, a train of bright solitons is excited. Here
the atoms in a bright soliton and in the background in a period
$[0,\Gamma]$ are $\int_{0}^{\Gamma} |\psi_{12}|^2dx\approx 3510$,
$\int_{0}^{\Gamma} |A_c\Omega|^2dx\approx 6815$, respectively.
Thus we can conclude that an important condition for exciting a
train of bright solitons is that the background is strong enough.

\subsection{\bf Dynamics of a Dark Soliton in BECs}

\begin{figure}[!ht] \centering
\renewcommand{\figurename}{Fig.}
\includegraphics[height=5cm,width=7cm]{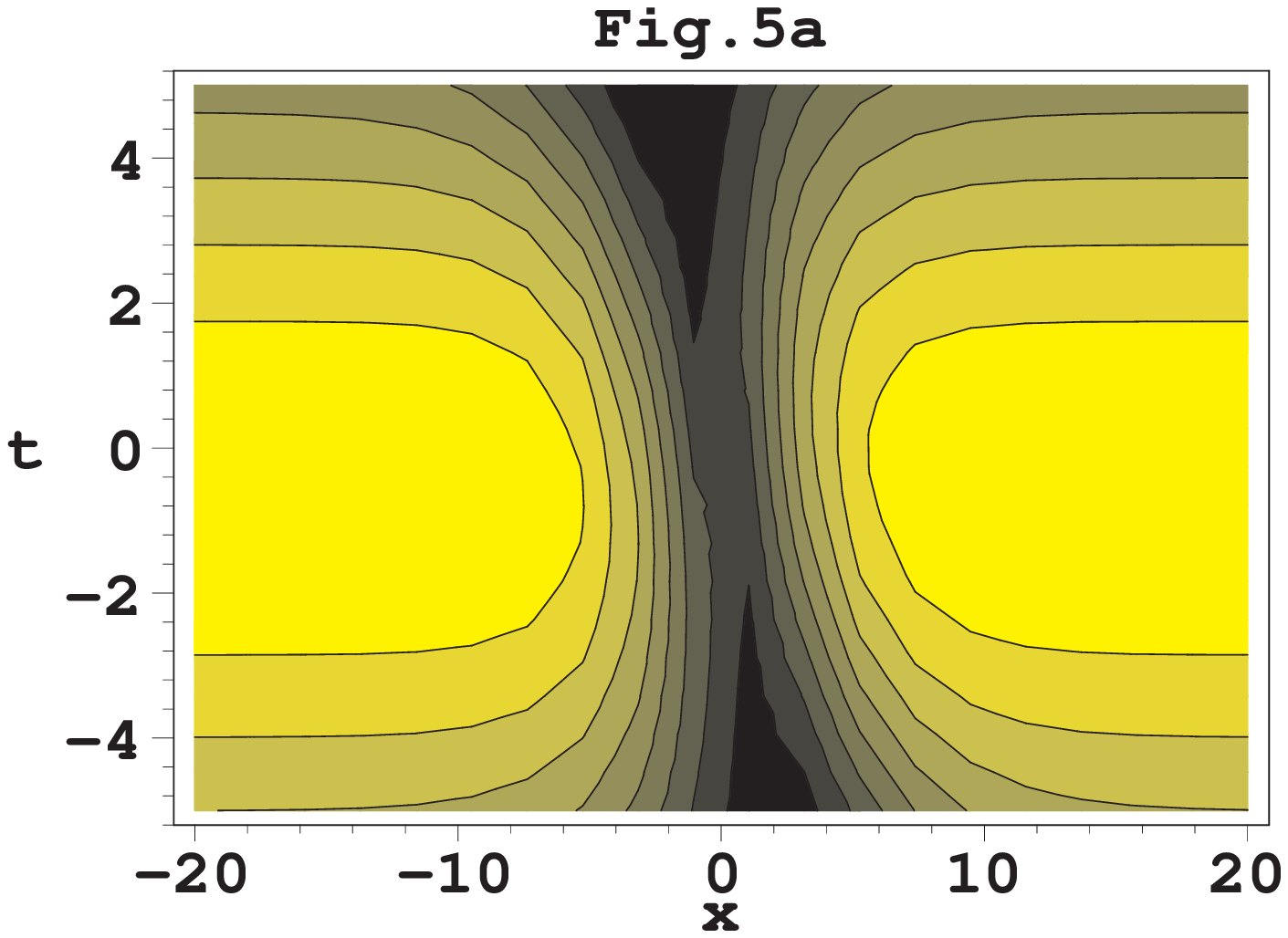}%
\includegraphics[height=5cm,width=7cm]{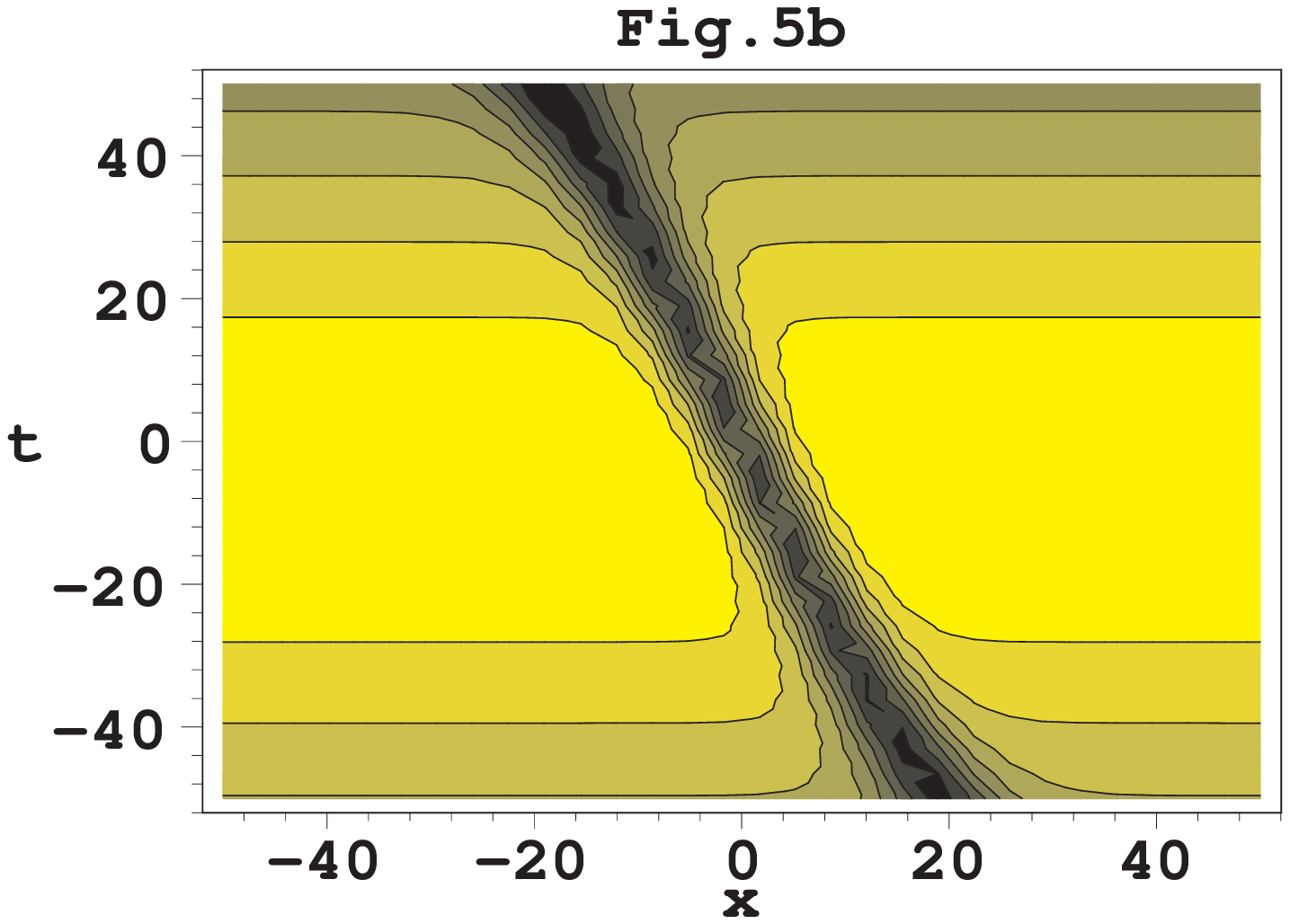}%
\caption{\small (Color online) The evolution contour plots of
$|\psi_2|^2$, where $A_c=7, A_s=12, g_0=0.4, k_1=-4.4$,
$\lambda=-0.197 $ and $\gamma=-0.01$ in Fig.5a ; $\lambda=-0.02$
$\gamma=-0.001$ in Fig.5b.}
\end{figure}

When $\lambda\rightarrow 0$ and $\gamma\rightarrow 0$, (\ref{c04})
is reduced to the dark soliton \cite{prl93(2004)240403},
\begin{equation}\label{c15}
 \psi_2 = \frac{\sqrt{{2}}}{2}[A_c+iA_s\tanh(\frac{\sqrt{g_0}}{2}(x+A_s(
k_1+\sqrt{g_0}A_c))t)]  \exp(i\Delta).
\end{equation}
Therefore, the solution (\ref{c04}) should be a time-dependent dark
soliton, which can be shown by Fig. 5. From (\ref{c04}), we can
obtain the intensities of the background as follows
\begin{equation}\label{c16}
    |\psi_c|^2=(A_c^2+A_s^2)\frac{{\rm sech}(\lambda t)}{2}\exp(2\gamma t).
\end{equation}
Therefore from (\ref{c16}), we guess that the solution (\ref{c04})
may describe an interesting physical process: there are "moving
stop", which may be realized by use of laser,  at both ends of the
cigar-axis direction.

Proceeding as the case of the bright soliton, we obtain
\begin{equation}\label{}
    \int_{-\infty}^{+\infty}|\psi_{2}|^2-|\psi_2(\pm \infty,t)|^2{\rm d}
    x=-\frac{2A_s^2}{\sqrt{g_0}}\exp(2\gamma t),
\end{equation}
which describes the region of decreased density and contains a
negative "number of atoms".

As shown in Fig.5, when the absolute of  $\lambda$ and $\gamma$ are
smaller, the dark solitons can keep a longer time and propagate a
longer distance. Under the conditions in Fig.5, the scattering
lengths are in a range $0.05{\rm nm}\leq a(t)\leq 0.20{\rm nm}$.
Therefore in order to keep a dark soliton a long time in BECs, we
should also reduce the values of $\lambda$ by adjusting the harmonic
oscillator frequencies $\omega_{\perp}$ and $\omega_0$ and reduce
the absolute value of $\gamma$ by controlling the loss of atoms.

\section{Conclusions}

In summary, we present a direct method to obtain two families of
analytical solutions for the nonlinear Schrodinger equation which
describe the dynamics of solitons in Bose-Einstein condensates
with the time-dependent interatomic interaction in an expulsive
parabolic and complex potential. The dynamics of a bright soliton,
a train of bright solitons and a dark soliton are analyzed
thoroughly. We can extend the lifetime of a bright soliton or a
dark soliton in BEC by reducing the ratio of the axial oscillation
frequency to radial oscillation frequency and control the loss of
atoms. Meanwhile, our results also demonstrate that a train of
bright solitons in BEC may be excited with a strong enough
background. It is very interesting to find these new phenomena
which are of special importance in the field of an atom laser in
further experiments.

B. Li would express his sincerely thanks to Profs. G. X. Huang and
Z. D. Li for their helpful discussions. This work is supported by
the NSF of China under Grant Nos. 10747141, 10735030, 90406017,
60525417, 10740420252, the NKBRSF of China under Grant 2005CB724508,
2006CB921400, Zhejiang Provincial NSF of China under Grant Nos.
605408, Ningbo NSF under Grant Nos. 2007A610049, 2006A610093 and
K.C.Wong Magna Fund in Ningbo University.

\end{document}